\documentclass[times,longtitle,nopreprintline]{elsarticle}
\usepackage{medima}

\usepackage{amssymb}
\usepackage{latexsym}
\usepackage{url}
\usepackage{xcolor}
\usepackage{tabu}
\usepackage{makeidx} 
\usepackage{times}
\usepackage{amsfonts}
\usepackage{color}
\usepackage{listings}
\usepackage{tikz}
\usepackage{ifthen}
\usepackage{multirow}
\usepackage{booktabs}
\usepackage{xspace}
\usepackage{mwe} 
\usepackage{framed}

\usepackage[colorlinks=true]{hyperref}
\usepackage{algorithm}
\usepackage{algpseudocode}
\usepackage{amsmath}
\usepackage{bbm}
\usepackage{multirow}

\graphicspath{figs/}

\begin{document}

% \verso{Dong Yang, Ziyue Xu \textit{et~al.}}

\begin{frontmatter}
\title{Federated Semi-Supervised Learning for COVID Region Segmentation in Chest CT using Multi-National Data from China, Italy, Japan}
\tnotetext[tnote1]{NIH has a cooperative research and development agreement with NVIDIA that involves artificial intelligence and deep learning using medical imaging. This research was supported in part by the Center for Interventional Oncology, NIH Grant\#1ZIDBC011242 \&1ZIACL040015 and the Intramural Research Program of the NIH.}

\author[1]{Dong \snm{Yang}\fnref{fn1}}
\author[1]{Ziyue \snm{Xu}\fnref{fn1}}
\fntext[fn1]{Authors contributed equally}
\author[1]{Wenqi \snm{Li}}
\author[1]{Andriy \snm{Myronenko}}
\author[1]{Holger R. \snm{Roth}}
\author[2,3]{Stephanie \snm{Harmon}}
\author[4]{Sheng \snm{Xu}}
\author[2]{Baris \snm{Turkbey}}
\author[5]{Evrim \snm{Turkbey}}
\author[1]{Xiaosong \snm{Wang}}
\author[1]{Wentao \snm{Zhu}}
\author[6]{Gianpaolo \snm{Carrafiello}}
\author[7]{Francesca \snm{Patella}}
\author[7]{Maurizio \snm{Cariati}}
\author[8]{Hirofumi \snm{Obinata}}
\author[8]{Hitoshi \snm{Mori}}
\author[8]{Kaku \snm{Tamura}}
\author[9]{Peng \snm{An}}
\author[4]{Bradford J. \snm{Wood}}
\author[1]{Daguang \snm{Xu}\corref{cor1}}
\cortext[cor1]{Corresponding author: daguangx@nvidia.com}

\address[1]{Nvidia Corporation, 4500 East West Highway, Bethesda, Maryland 20814, USA}
\address[2]{Molecular Imaging Branch, National Cancer Institute, NIH, Bethesda, MD USA}
\address[3]{Frederick National Laboratory for Cancer Research, Leidos Biomedical Research, Inc., Molecular Imaging Branch, National Cancer Institute, NIH, Bethesda, MD USA}
\address[4]{Center for Interventional Oncology, Radiology and Imaging Sciences, NIH Clinical Center and National Cancer Institute, Center for Cancer Research, National Institutes of Health, Bethesda, MD USA}
\address[5]{Radiology and Imaging Sciences, NIH Clinical Center, National Institutes of Health, Bethesda, MD USA}
\address[6]{Radiology Department, Fondazione IRCCS Cà Granda Ospedale Maggiore Policlinico, University of Milan, Italy}
\address[7]{Diagnostic and Interventional Radiology Service, San Paolo Hospital; ASST Santi Paolo e Carlo; Milan, Italy}
\address[8]{Self-Defense Forces Central Hospital, Tokyo, Japan}
\address[9]{Department of Radiology, Xiangyang First People's Hospital Affiliated to Hubei University of Medicine Xiangyang, Hubei, China}

%\received{1 June 2020}
% \finalform{10 May 2013}
% \accepted{13 May 2013}
% \availableonline{15 May 2013}
% \communicated{}

\begin{abstract}
The recent outbreak of Coronavirus Disease 2019 (COVID-19) has led to urgent needs for reliable diagnosis and management of SARS-CoV-2 infection. The current guideline is using RT-PCR for testing. As a complimentary tool with diagnostic imaging, chest Computed Tomography (CT) has been shown to be able to reveal visual patterns characteristic for COVID-19, which has definite value at several stages during the disease course. To facilitate CT analysis, recent efforts have focused on computer-aided characterization and diagnosis with chest CT scan, which has shown promising results. However, domain shift of data across clinical data centers poses a serious challenge when deploying learning-based models. A common way to alleviate this issue is to fine-tune the model locally with the target domain’s local data and annotations. Unfortunately, the availability and quality of local annotations usually varies due to heterogeneity in equipment and distribution of medical resources across the globe. This impact may be pronounced in the detection of COVID-19, since the relevant patterns vary in size, shape, and texture. In this work, we attempt to find a solution for this challenge via federated and semi-supervised learning. A multi-national database consisting of 1704 scans from three countries is adopted to study the performance gap, when training a model with one dataset and applying it to another. Expert radiologists manually delineated 945 scans for COVID-19 findings. In handling the variability in both the data and annotations, a novel federated semi-supervised learning technique is proposed to fully utilize all available data (with or without annotations). Federated learning avoids the need for sensitive data-sharing, which makes it favorable for institutions and nations with strict regulatory policy on data privacy. Moreover, semi-supervision potentially reduces the annotation burden under a distributed setting. The proposed framework is shown to be effective compared to fully supervised scenarios with conventional data sharing instead of model weight sharing.
\end{abstract}


\end{frontmatter}

\section{Introduction}
\label{sec:intro}
\begin{figure*}[h]
\begin{center}
    \includegraphics[width=18.0cm]{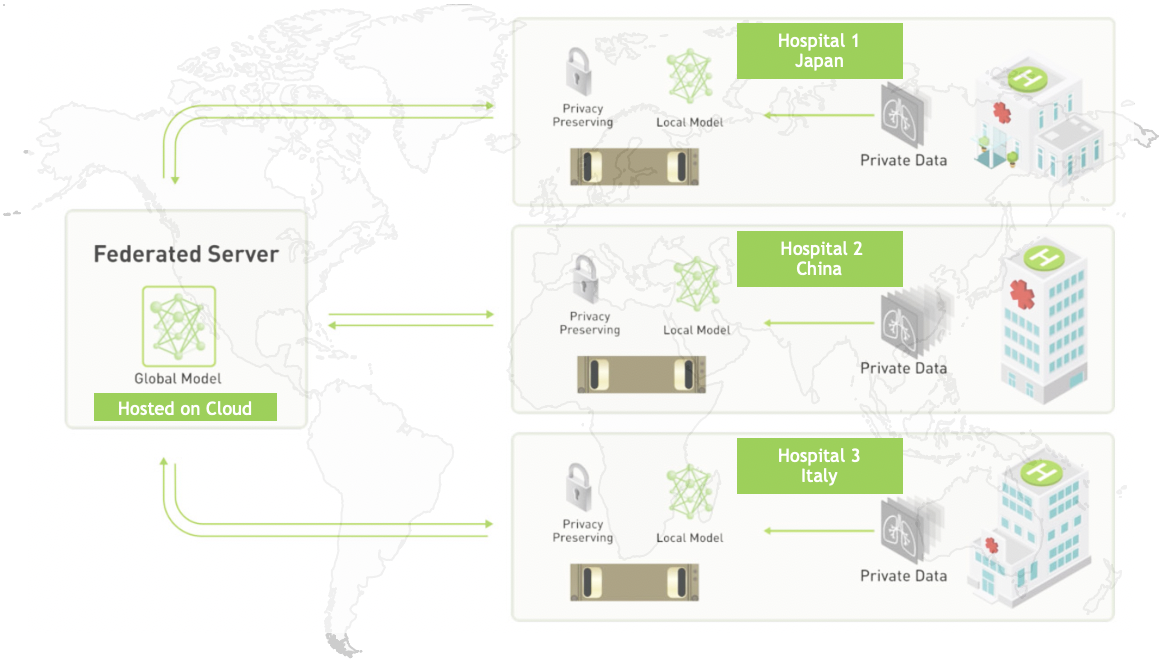}
\end{center}
%\vspace{-0.5cm}
\caption{
    Federated learning with privacy preserving in medical imaging. 
    %Although the central server communicates with clients from multi-national institution without any sensitive data exchanging, the global model benefits from weights and gradients from client's local models.
    The central federated server communicates with clients from multi-national institutions by sharing weights and gradients of models, without exchanging any sensitive data information.
}
\label{fig:fl}
\end{figure*}

The COVID-19 pandemic has caused millions of confirmed cases and hundreds of thousands of deaths globally. Early and effective diagnosis is among critical measurements to control the infectious disease and manage its spread. Current standard test for potential SARS-CoV-2 infection is via RT-PCR. However, this technique can have high false negative/positive rate for coronavirus due to multi-factorial issues~\citep{nejm2003sars}. 

As a complimentary approach, medical imaging, specifically chest CT, is undergoing investigations on its capability of revealing important characteristics of COVID-19. Although its specificity for diagnostic purposes lacks consensus and is not recommended according to the guidelines from the ACR~\citep{acr}, researchers have been looking for visual patterns related to viral infection from the images. A better understanding of the appearance of COVID-19 in chest CT may serve as an epidemiologic tool for mitigating outbreaks. Current consensus on the imaging pattern is the presence of infiltrates, from ground glass opacity to consolidation, usually bilateral, multilobar, with a peripheral distribution~\citep{nejm2020,radiology2020}.

Quantification and characterization over the region of infiltrates may further help the understanding and tracking of disease progression, and provide important information of this novel virus in an outbreak situation~\citep{ctquant2020}. In order to achieve this goal, detection and delineation of disease patterns are required. Unfortunately, such processes can be extremely tedious and time-consuming, especially at this time when many experts are already working around the clock on their clinical duty during a pandemic. Therefore, automated computerized methods are being developed and deployed to facilitate the understanding of the findings from medical images~\citep{ieeereview2020}. Among them, methods based on deep learning techniques often achieve the state-of-the-art accuracy~\citep{radiologyai,tmi1,tmi2,tmi3,tmi4}.  

To design and develop an artificial intelligence (AI) system that can properly and robustly handle this problem, large amounts of data, as well as annotations, from diverse sources are required. However, in reality: 1) data access is often limited by strict sharing policies on sensitive private patient information, and 2) annotations have great variances in both quantity and quality due to experts' experience, availability and cost across imaging sites. Hence, most existing methods are trained using limited amount of data from a single site. On the other hand for model deployment, a well-known challenge for deep learning is the ``domain shift'' caused by the distribution difference between source data and target data, often leading to significant performance variance or degradation among different sites. Therefore, a mechanism enabling cross-institution collaboration under the constraint of significant difference in annotation availability and strict data sharing policies is highly desirable and may facilitate AI model development for COVID-19.

In this work, we propose a novel system to address the aforementioned challenges, which is based on federated and semi-supervised learning. Federated learning~\citep{li2020federated}, especially ones with secure features~\citep{li2019privacy}, can often give sufficient flexibility to different institutions to collaboratively train deep learning models without data sharing, as shown in Fig.~\ref{fig:fl}; while semi-supervised learning can ensure effective training even when some sites have only limited amount of annotated data but large amount of unannotated data. In addition, the semi-supervised setting could reduce the burden of experts annotation, which is very valuable in current pandemic situation.

To test the proposed framework, we chose the task of segmentation for abnormal regions related to COVID-19, which is the most time-consuming as compared with other tasks like classification. This is because for most cases, slice-by-slice delineation is needed. With various configurations of the proposed framework, we show that our method's design can naturally benefit from multiple heterogeneous data sources under semi-supervise scenario. Finally, our method is task and training pipeline independent which makes it easy to be adapted to other deep learning tasks, such as COVID classification in CXR/CT, and non-COVID imaging tasks.

\section{Related Work}
\label{sec:related_work}
\noindent\textbf{Federated Learning} is an advanced distributed learning concept that takes advantage of datasets across multiple institutions without any explicit training data centralization or sharing~\citep{yang2019federated,li2020federated}.
Although federated learning (FL) was initially designed for mobile edge devices, it has attracted increasing attention in healthcare domain because of its privacy preserving nature of the patient information. FL is agnostic to the type of the input data. It is capable of analyzing various medical data modalities, from free-text clinical reports to high-dimensional medical images~\citep{xu2019federated}.
\citet{brisimi2018federated} adopted FL to train a predictive model and solve a support vector machine problem for analysis of electronic health record (EHR) data.
FL was applied for wearable healthcare using personalized machine learning models~\citep{chen2020fedhealth}.
\citet{li2020federated} built an FL framework for multi-site fMRI classification with preserved privacy.
Recently, FL has been successfully applied on multi-institutional brain MRI for tumour segmentation with deep neural networks and improved privacy preserving of patient information~\citep{sheller2018multi,li2019privacy}.

\noindent\textbf{Semi-Supervised Learning} leverages the available information of unlabeled data, together with the supervision from labeled data, to improve the effectiveness and generalizability of machine learning models.
In computer vision, semi-supervised learning has been investigated from different perspectives for various applications (e.g. image recognition).
To take advantage of unlabeled data, consistency constraints have been investigated to mitigate the gap within and between domains of labeled and unlabeled data~\citep{verma2019interpolation,berthelot2019mixmatch,berthelot2019remixmatch,sohn2020fixmatch,liu2020semi}.
One research trend is the framework of teacher-student models, which perform well utilizing consistency constraints between models for labeled and unlabeled data, respectively~\citep{tarvainen2017mean,luo2018smooth}.
Another similar framework, ``noisy-student'', achieved the state-of-the-art performance on ImageNet classification when jointly trained with a large amount of unlabeled data~\citep{xie2019self}.
Meanwhile, consistency-based model regularization can be implemented using model predictions of unlabeled data with data augmentation or pre-processing ~\citep{berthelot2019mixmatch,berthelot2019remixmatch,verma2019interpolation,sohn2020fixmatch}.
Another trend is to design auxiliary supervised tasks for unlabeled data, such as solving jigsaw puzzles~\citep{noroozi2016unsupervised}, predicting rotation angles~\citep{gidaris2018unsupervised,zhai2019s4l}. Alternatively, co-training \citep{qiao2018deep} has been applied to semi-supervised image recognition where different models are trained on different ``views'' in order to learn complimentary information from the data.
In general, most studies in the field focus on large-scale image recognition using 2D convolutional neural networks.
Some works covered semantic segmentation~\citep{hong2015decoupled,papandreou2015weakly}, object detection~\citep{misra2015watch,tang2016large} in 2D, as well as in graph-structured data analysis~\citep{kipf2016semi}.

\noindent\textbf{Semi-Supervised Learning in Medical Imaging}~becomes a popular topic as large-scale datasets are made publicly available (e.g.~\cite{cheplygina2018not}).
But in the meantime, it is difficult to collect annotation for all datasets from experts or radiologists in practice.
\cite{bai2017semi} introduced a semi-supervised learning methods for cardiac image segmentation, using both deep neural networks and conditional random field (CRF).
\cite{li2018semi} proposed to add model regularization based on image rotation and flipping for unlabeled data in skin lesion segmentation. 

Several works used self-ensembling and teacher-student interaction for medical image analysis, e.g.~\cite{liu2020semi}.
~\cite{cui2019semi} proposed an adapted mean teacher model to improve accuracy of brain lesion segmentation leveraging both annotated and unannotated data.~\cite{yu2019uncertainty} added an uncertainty-aware scheme to the teacher-student framework improving consistency regularization at training for left atrium segmentation in 3D MRI.

The multi-plane information from 3D medical images was used to enhance the supervised training model for prediction consistency in a co-training approach~\citep{zhou2018semi}. This method was further extended leveraging multi-view information of 3D medical images for full volumetric segmentation~\citep{xia20203d}.
Moreover, some researchers studied the usage of mixed supervision for medical image analysis~\citep{shah2018ms,mlynarski2019deep}.

\noindent\textbf{Federated Semi-Supervised Learning}~brings up challenges and complexity about how to exploit unlabeled data under a distributed learning setting~\citep{jin2020survey}.
Unsupervised federated learning has been investigated for representation learning in a distributed setting~\citep{van2020towards}.
Federated self-learning was shown to be capable of detecting abnormality without any data label~\citep{nguyen2019diot}.
Moreover, one-shot federated learning was introduced to conduct single-round communication between clients and server for both supervised and semi-supervised learning~\citep{guha2019one}.

It is important to note that, in the context of federated learning, ``semi-supervised learning'' has a new dimension beyond the centralized training scenario, such as teacher-student model or co-training methods listed above. This new dimension is introduced by the multi-client setting and that different clients can have completely different annotation availability. We call it ``global semi-supervision'', as compared with ``local semi-supervision'' that is potentially viable for each individual client. As a matter of fact, the client-level network is compatible with other semi-supervised techniques, such as teacher-student model. In this paper, we propose an efficient and robust solution for the ``global semi-supervision’’ problem. To the best of our knowledge, there are very limited existing works on such federated semi-supervised learning or federated self-learning on unlabeled medical imaging data.
In this paper, we propose a federated semi-supervised learning framework which utilizes unlabeled data to improve FL training and validate it in COVID region segmentation task. Our framework enables cross-institutional training on large-scale heterogeneous datasets without sharing sensitive private information.  
%, given the advantages and complexities from both federated learning and semi-supervised learning.
%
\section{Coronavirus Affected Region Segmentation}
\label{sec:seg}
\begin{figure}[h]
\begin{center}
    \includegraphics[width=9cm]{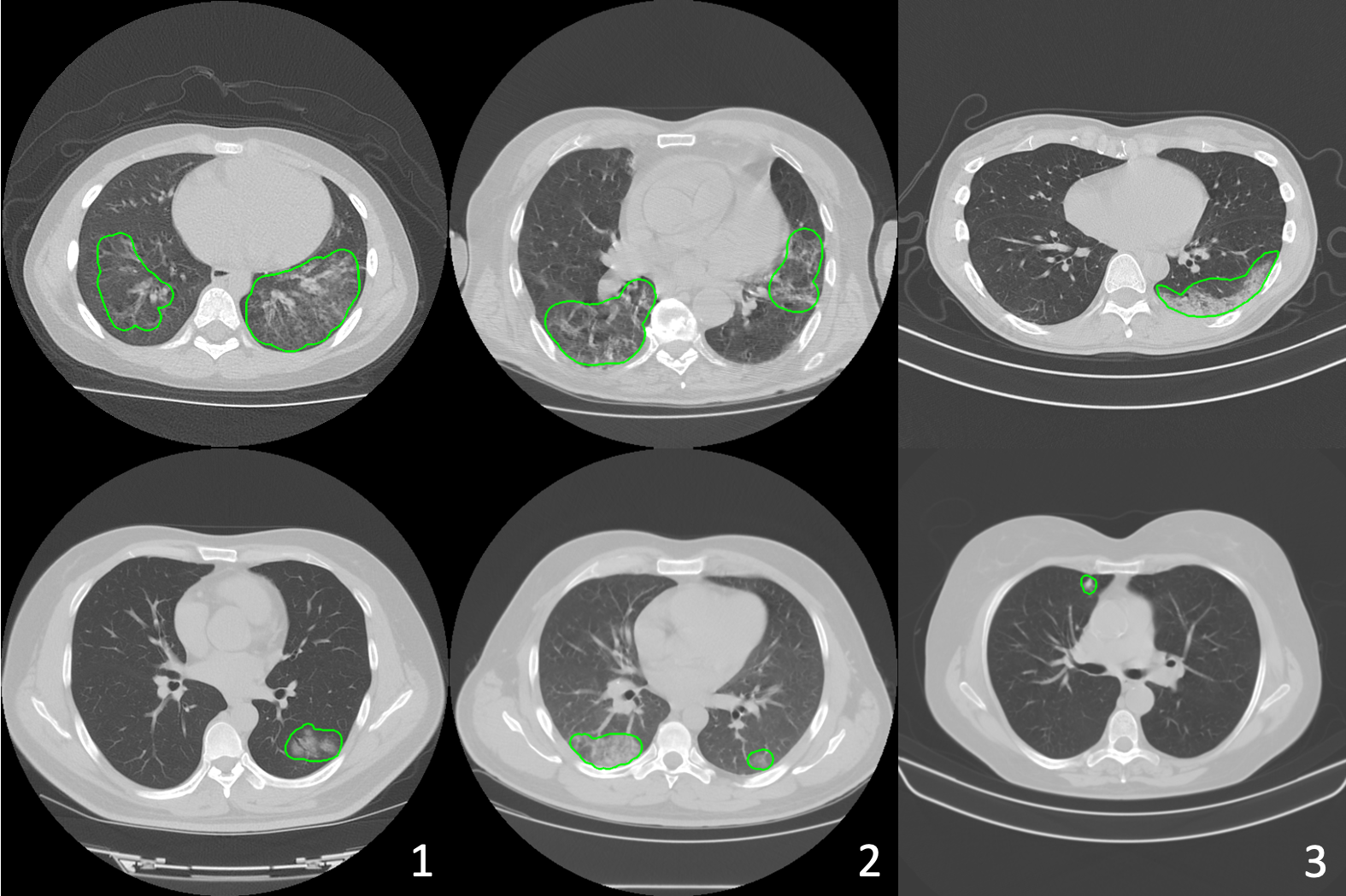}
\end{center}
\caption{
    The axial planes of chest CT scans from three different sites. Areas inside green contours represent COVID-19 affected regions annotated by radiologist. The appearance of the affected region identified as ``infiltrates'' range from diffused ground glass opacity (COVID, upper row) to focal nodules (lower row).
}
\label{fig:ct}
\end{figure}

\begin{figure}[h]
\begin{center}
    \includegraphics[width=9cm]{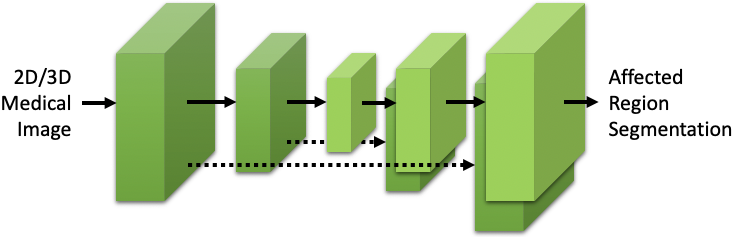}
\end{center}
\caption{
    The general architecture of fully convolutional network (FCN) for COVID-19 affected region in medical imaging. Dashed lines denote skip connections which feed earlier feature maps to later neural network layers.
}
\label{fig:network}
\end{figure}

Nowadays, machine learning based methods have been developed for medical imaging data acquisition, segmentation, and diagnosis of COVID-19~\citep{dong2020role,shi2020review}.
Relying on the success of deep learning in medical image analysis, imaging characteristics of COVID-19 have been studied and analyzed from various perspectives.
Some examples of affected regions of COVID-19 is given in Fig.~\ref{fig:ct}. For the identification and segmentation of such regions, there are two major research directions using deep neural networks.
The first one is to use classification models to distinguish normal subjects and patients.
Class activation maps (CAM) can be extracted from these models that correspond to the affected region inside the lung area~\citep{bai2020ai,li2020artificial,mei2020artificial,wang2020prior}.
The second direction is to apply 3D segmentation networks, typically fully convolution networks (FCN), and directly extract the COVID-19 affected regions following an image-to-image fashion~\citep{fan2020inf,huang2020serial,liu20203d,shan2020lung,xie2020contextual,zhang2020clinically,zhou2020automatic} shown in Fig.~\ref{fig:network}.

In this paper, we follow the second approach, using 3D U-shape FCN~\citep{liu20183d} as our baseline model, to segment the ground glass-like opaque (GGO) regions (COVID-19 affected regions, or COVID for short) of the lung from the 3D chest CT. However, it is noteworthy that our framework is independent of the exact neural network architecture used.
This 3D U-shape FCN can be initialized with 2D model. Specifically, during the model initialization, weights of the 3D encoder are directly transferred from the pre-trained 2D ResNet with necessary operation conversion. For instance, 2D 3$\times$3 convolutions are converted to 3D 3$\times$3$\times$1 convolutions with the same amount of parameters. The parameters of batch normalization can be transferred directly without any modification.
%The segmentation of GGO regions is formulated as the voxel-wise binary classification.
Each voxel is predicted as either foreground (COVID) or background.
The output of models is a two-channel probability map after soft-max activation.
The model is partially initialized with pre-trained weights of ResNet~\citep{he2016deep} from ImageNet classification for the encoder layers. We use the Adam optimizer to minimize the soft Dice loss~\citep{milletari2016v} given as
\begin{equation}
    \mathcal{L}_{dice} = 1 - \frac{2\sum_{i}y_i\hat{y}_i}{\sum_{i} (y_i)^2 + \sum_{i} (\hat{y}_i)^2}.
    \label{eq:dice_loss}
\end{equation}
Here, $y$ represents ground truth label and $\hat{y}$ is the prediction from the deep learning model.

\noindent\textbf{Implementation Details.}~The CT volumes are converted to the most frequent resolution $0.8\mathrm{mm} \times 0.8\mathrm{mm} \times 5.0\mathrm{mm}$ of the chest CT datasets before training and testing/inference.
The actual input of the model is a cropped region-of-interest (ROI) with fixed size of $160 \times 160 \times 32$ during training. 
The patches sampled from the CT volumes are fed into the network for training. Patches are sampled with equal chance from foreground or background regions to maintain the training sample balance.
The intensity of CT is clipped between Hounsfield units (HU) $0$ and $-1000$, and mapped to the range of $\left [ 0,1 \right ]$.
The data augmentation strategy includes random flipping, random rotation, and random intensity shift.
Moreover, the deep learning model is trained using 16GB NVIDIA Tesla V100 GPUs, with the pipeline developed using NVIDIA Clara Train SDK Platform~\citep{clara2020}.
To achieve segmentation of the entire CT at inference, we use a crop size of $224 \times 224 \times 32$ to increase inference speed and to avoid artifacts caused by cropping. The sliding step is 16 along three axes.
The batch size for training is 4, and learning rate is set to $0.0001$.
\section{Federated Semi-Supervised Learning}
In this section, we introduce our framework, federated semi-supervised learning, for COVID region segmentation in 3D chest CT.
The framework is designed to leverage unlabeled data for federated learning.
In the following sub-sections, we explain the mechanism of our frameworks in details: the fundamentals of federated learning for COVID region segmentation are introduced in Sec.~\ref{sub:fl}; then we illustrate the settings of federated semi-supervised learning in the segmentation task in Sec.~\ref{sub:flsemi}; last but not least, the detailed implementation is further discussed in Sec.~\ref{sub:imple}.

\begin{algorithm}[h]
\caption{Federated learning for COVID region segmentation using weighted federated averaging.}
\textbf{Input:}~number of clients $\mathcal{C}$, amount of global synchronization rounds $\mathcal{T}$, aggregation weights $\mathcal{W}$ for all clients, learning rate $\lambda_i$ for each client $i$ (for simplicity, we show the gradient descent update rule; not Adam optimization).\\
\textbf{Output:}~optimal $\theta_c$ for each client $c$.
\begin{algorithmic}[1]
    \Procedure{ServerUpdate}{}
        \State{initialize global model $\theta_0$}
        \State{send $\theta_0$ to all $\mathcal{C}$ clients}
        \For{round $t=1,2,\cdots,\mathcal{T}$}
            \State{wait for all $\mathcal{C}$ to finish update}
            \State{$\left (\Delta_\theta^i,n_i  \right )\leftarrow\texttt{ClientUpdate}\left (\theta_{t-1}\right )$, $i=1,\cdots,\mathcal{C}$}
            \State{$\hat{w}_i\leftarrow \left( n_i / \sum_i{n_i}\right)\cdot w_i$, $w_i \in \mathcal{W},i=1,\cdots,\mathcal{C}$}
            \State{$\hat{\Delta}_{\theta,t-1} \leftarrow \sum_{i=1}^{\mathcal{C}}\hat{w}_i\cdot\Delta_\theta^i$}\Comment{aggregate updates}
            \State{$\theta_t\leftarrow\theta_{t-1} + \hat{\Delta}_{\theta,t-1}$}
        \EndFor
    \EndProcedure
    \item[]
    \Function{ClientUpdate}{$\theta$}
        \State{$n\leftarrow$ amount of total training iterations}
        \State{$\theta_{\mathrm{init}}\leftarrow\theta$}
        \For{iteration $j \in 1,2,\cdots,n$}
            \State{$\theta\leftarrow\theta-\lambda_i\triangledown \mathcal{L}\left ( \theta,j \right )$}\Comment{optimize loss function}
        \EndFor
        \State{$\Delta_\theta\leftarrow\theta-\theta_{\mathrm{init}}$}
        \State\Return{$\left (\Delta_\theta,n  \right )$}
    \EndFunction
\end{algorithmic}
\label{alg:fl}
\end{algorithm}
\subsection{Federated Learning}
\label{sub:fl}
In our paper, the federated learning framework follows the conventional settings as~\citep{mcmahan2016communication,li2019privacy}.
In the setting, a single server hosts the global optimal model at the moment, and meanwhile communicates with multiple clients.
The neural network architecture is shared between server and clients.
The communication is only about weight or gradient transferring, which is synchronized for all clients.
\begin{figure}[h]
\begin{center}
    \includegraphics[width=7.5cm]{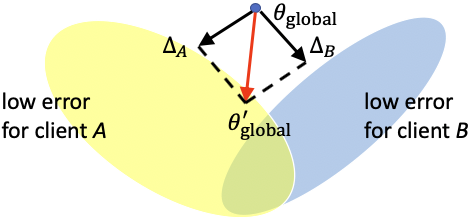}
\end{center}
\caption{
    FL is to find global low-error space of $\theta_\mathrm{global}$ for client $A$ and $B$ after aggregation of ``gradients'' $\Delta_A$ and $\Delta_B$. The low-error space is defined by each client, which could correspond to high accuracy, or high model consistency for self-supervision.
}
\label{fig:vis}
\end{figure}
All the data is associated with the clients.
The data point from one client is invisible to both the server and other clients.
There is usually no complicated computation on server side other than simple weighted aggregation.
During FL training, the server collects ``gradients'' $\Delta_\theta$ from clients simultaneously, aggregate those gradients, and send new model weights back to clients.
The most important job for the server is \textbf{weight aggregation}.
As shown in Eq.~\ref{eq:agg}, the aggregation is conducted as weighted summation.
\begin{equation}
    \hat{\Delta}_{\theta} \leftarrow \hat{w}_i\cdot\sum_{i=1}^{\mathcal{C}}\Delta_\theta^i
    \label{eq:agg}
\end{equation}
Here, $C$ is the number of clients. We firstly weigh clients by $n$, the quantity of iterations per synchronization round.
Because the model trained with more steps tends to have larger difference with the initial model,
the updates from clients should be normalized to have the same update pace, which would avoid the unnecessary bias towards any one of the clients.
Therefore, besides weights from iterations, we have additional weights $\mathcal{W}$ for each client assigned by users.
Hence, the overall weight of each client contains two components as follows.
\begin{equation}
    \hat{w}_i\leftarrow \frac{n_i}{\sum_i{n_i}}\cdot w_i,w_i \in \mathcal{W},i=1,\cdots,\mathcal{C}
\end{equation}
Here, $n_i$ is the iteration number per round of client $i$.
Once aggregation is accomplished, the server sends out the updated model weights back to each client at the same time.

Next, the clients collect models weights from the server, finetune the model with their local data, and send out the new gradient to server.
They are independent instances, which are not direction connection between each other.
%Thus, the communication between client and server is straightforward.
%Besides communication, the client focus on training models using the data at the client side.
%, and pre-trained model from server.
In this paper, the clients launch training jobs described in Sec.~\ref{sec:seg} for COVID region segmentation.
Each client has its own chest CT data, and GPUs as computing resource.
Moreover, the optimal model checkpoint for each client is selected based on its own validation set.

The server is launched first and the clients are initialized accordingly with the global model from server.
After several FL rounds of server-client communication, the global model on the server would be improved with greater generalizibility, and the performance of local models would be further boosted.
This intuition is shown in Fig.~\ref{fig:vis}.
The entire FL algorithm is shown in Alg.~\ref{alg:fl} and is potentially feasible for all deep learning related applications of medical image analysis.
\subsection{Federated Semi-Supervised Learning on COVID region segmentation}
\label{sub:flsemi}
We assume that some of clients may not have enough expertise to create accurate annotation for their datasets.
But the information of their patient data is still valuable to the rest of community.
In practice, some clients only possess unlabeled chest CT images, which could be from COVID-19 patients, pneumonia patients, or normal subjects.
However, each of those database has its own merits for jointly training COVID-19 affected region segmentation.
For instance, COVID-19 related CT images contain informative context from its appearance inside the lung region.
Non COVID-19 related CT images can provide guidance on false positive removal for the COVID-19 affected region segmentation model (ideally nothing should be predicted based on those CT images).
In order to  utilize the clients' side unlabeled data, we propose a novel framework of federated semi-supervised learning for COVID-19 affected region segmentation.

\begin{algorithm}[h]
\caption{Self-supervised learning algorithm at clients with unlabeled data for COVID region segmentation.}
\textbf{Input:}~learning rate $\lambda$ for each client, unlabeled data pool $\mathcal{U}$, global model $\mathcal{H}$.\\
\textbf{Output:}~weight difference $\Delta_\theta$.
\begin{algorithmic}[1]
    \Function{SelfSupervisedClientUpdate}{$\theta$}
        \State{$n\leftarrow$ amount of total training iterations}
        \State{$\theta_{\mathrm{init}}\leftarrow\theta$}
        \For{iteration $j \in 1,2,\cdots,n$}
            \State{$\hat{u}_j\leftarrow \mathrm{Augment}\left(u_j\right)$, $u_j\in\mathcal{U}$}\Comment{input perturbation}
            \State{$\mathcal{L}\left ( \theta,j \right )\leftarrow\mathcal{L}_\mathrm{consistency}\left(\mathcal{H}\left(u_j;\theta\right),\mathcal{H}\left(\hat{u}_j;\theta\right)\right)$}
            \State{$\theta\leftarrow\theta-\lambda\triangledown \mathcal{L}\left ( \theta,j \right )$}\Comment{optimize loss function}
        \EndFor
        \State{$\Delta_\theta\leftarrow\theta-\theta_{\mathrm{init}}$}
        \State\Return{$\left (\Delta_\theta,n  \right )$}
    \EndFunction
\end{algorithmic}
\label{alg:semi}
\end{algorithm}

Under the existing FL setting, we create a unsupervised client, which only has unlabeled data, without modifying the FL server.
The unsupervised clients retrieve the global model $\mathcal{H}\left(\cdot;\theta\right)$ with weights $\theta$ from the server and apply it to their own database.
Then, we would like to enforce consistency of predictions from the global model, to further adjust model weights.
Similar to~\citep{berthelot2019mixmatch,berthelot2019remixmatch,sohn2020fixmatch}, we introduce a new loss function based on data augmentation.
The assumption behind this is that the generalizable model should perform the same with original data and slightly perturbed data.
We assume that the perturbation can still be within the range of the actual data distribution.

Let $u$ denote one CT image from the unlabeled database $\mathcal{U}$.
We create the pseudo-label $\bar{y}$ using prediction of the current model with input $u$ and hard thresholding (against 0.5).
\begin{equation}
  \bar{y} =
    \begin{cases}
      1, & \mathcal{H}\left(u\right) > 0.5\\
      0, & \text{otherwise}
    \end{cases}
\end{equation}
In order to train the global model locally, the new loss function minimizes the difference between pseudo-label $\bar{y}$ and the prediction after augmentation $\emph{g}\left(\cdot\right)$:
\begin{equation}
    \mathcal{L}_\mathrm{consistency}=\mathcal{L}\left(\bar{y},\mathcal{H}\left(\emph{g}\left(u\right)\right)\right).
\end{equation}
Here, $\mathcal{L}$ is the soft Dice loss for segmentation tasks, as in Eq.\ref{eq:dice_loss}.
Other loss functions, such as cross entropy or $\ell_2$ loss, can also be used here.
Data augmentation could be random scale shift, Cutout~\citep{devries2017improved}, adding Gaussian noise, and so on.
In general, $\emph{g}\left(\cdot\right)$ on image appearance should not generate out-of-distribution samples.
Because the pseudo-label is generated fully from the image, the ambiguous area (output probability close to $0.5$) in the prediction may not be helpful for model training.
To further improve the quality of pseudo-label, a confidence threshold $\tau\in\left[0.5,1\right]$ can be added to determined the regions used for computing the loss:
\begin{equation}
    \mathcal{L}_\mathrm{consistency}=\mathbbm{1}\left(\mathcal{H}\left(u\right)>\tau\right)\mathcal{L}\left(\hat{y},\mathcal{H}\left(\emph{g}\left(u\right)\right)\right).
\end{equation}

\noindent\textbf{Comparison with Centralized Semi-Supervised Learning.}~After training several epochs, the server collects gradients $\Delta_\theta$ from both supervised clients $\mathcal{C}_s$ and self-supervised/unsupervised clients $\mathcal{C}_u$ for model aggregation:
\begin{equation}
    \hat{\Delta}_{\theta} \leftarrow \sum_{i\in\mathcal{C}_s}\hat{w}_i\cdot\Delta_\theta^i + \sum_{i\in\mathcal{C}_u}\hat{w}_i\cdot\Delta_\theta^i.
\end{equation}
Although it seems to be similar with jointly loss training in centralized semi-supervised learning (SSL), the proposed FL approach has intrinsic differences compared to SSL.
First, the objective function $\mathcal{L}$ is only visible to each client itself.
The gradients of clients may diverge to different local minima, which is also known as weight divergence problem in FL.
Second, for the communication efficiency of FL training, the model update cannot be conducted per local iteration.
It makes the federated semi-supervised learning even more challenging.
The weights $\hat{w}_i$ between different clients, and the learning rate $\lambda_i$ of each clients may play an important role for the final model performance.
Third, due to the difference of patients population, scanning protocol, and scanners, the data distribution of clients are usually non-identical and non-independent (non-iid) in the application.
%Although the data distributions of clients are non-identical and non-independent (non-iid) in the application, but the domains may not be largely overlapped.
For instance, one client has data from patients at early stage, and another client may possess data with severe conditions only.
There is clear appearance/domain difference between these two clients.
It is unclear how to handle such domain difference under federated semi-supervised learning setting.
All these factors make federated semi-supervised learning much more difficult comparing to centralized semi-supervised learning.

The federated semi-supervised algorithm for clients with unlabeled data is shown in Alg.~\ref{alg:semi}, which is also feasible for all machine learning related applications of medical image analysis.
%Under current setting, there cannot be unsupervised clients only without any labeled data.
Since the segmentation model requires necessary supervision to train, at least one client needs to possess labeled data in training.
However, federated unsupervised learning would be a future direction to explore for representation learning.
\subsection{Implementation Details}
\label{sub:imple}
The FL implementation for COVID-19 affected region segmentation is constructed with NVIDIA Clara Train SDK using TensorFlow 1.14~\citep{clara2020}, which uses the gPRC protocol for communication between the server and clients during model training.
Since we use patch-based training strategy for 3D images (because of GPU memory limited and efficient training), we modify the client training with the same iterations per epoch, and the same epochs per round.
Thus, the contributions from different clients are equivalent if the clients' weights are the same.

Training-from-scratch for unsupervised clients are not meaningful because no guidance from the global model or local unlabeled data is available.
This issue may be mitigated after several rounds of model aggregation, but it slows down the overall FL training efficiency.
Therefore, to enable reasonable training of the unsupervised client, for the federated learning with 1 supervised and 1 unsupervised clients, we first train the supervised client for 500 epochs (on the data of the supervised client). The federated learning then starts with this model and gets trained for another 500 epochs, such that the unsupervised client can start with a meaningful feature representation. For cases with two supervised clients, the federated learning starts from scratch and gets trained for 1000 epochs.

Each epoch contains 20 iterations, and each round of synchronization contains 20 epochs.
To be specific, we set the same iteration numbers per federated round for all clients to ensure the similar training paces of clients and mitigate potential side-effects caused by client asynchronization. Since the clients with more training iterations (or optimization steps) tend to converge faster than the one with less iterations.
Validation is conducted also every 20 epochs for supervised clients.
Then the optimal model checkpoint is determined using the validation dataset on supervised clients.

Moreover, the learning rate is $5e^{-6}$ for unsupervised clients.
It cannot be as large as the one in supervised clients because large learning rate could cause the model to overfit the self-supervised tasks quickly, potentially diverging the gradient towards a biased direction.
The same assumption can be made for client weights.
The weights of unsupervised clients cannot be larger than ones in supervised client, otherwise the convergence of global segmentation model becomes slow and unstable.
The actual input of the model is a cropped region-of-interest (ROI) with fixed size $160 \times 160 \times 32$ for training.
The patches sampled from the CT volumes are fed into the network for training, and they are sampled randomly over the entire CT volume.
And the intensity of CT is clipped between Hounsfield units (HU) $0$ and $-1000$, and mapped to the range $\left [ 0,1 \right ]$.
Adam optimizer is used to minimize consistency loss.
\section{Experiment and Results}
\label{sec:result}
\subsection{Data and Expert Annotation}
\noindent\textbf{COVID-19 Population:} patients undergoing CT evaluation with SARS-CoV-2 infection confirmed by RT-PCR were identified from three international centers: 1) 736 scans of 700 patients from the First Affiliated Hospital of Hubei University of Medicine in Hubei Province, China (referred to as \textbf{Image\_1}), 2) 496 scans of 244 patients from the Self-Defense Forces Central Hospital, Tokyo, Japan (referred to as \textbf{Image\_2}), and 3) 472 scans of 147 patients from San Paolo Hospital, Milan, Italy (referred to as \textbf{Image\_3}). It is important to note that the image acquisition from these three institutions varies considerably: in China, the CT scans were routinely obtained on the same day as a positive RT-PCR in an acute setting during the initial outbreak period; in Japan, the patients were a mixture of incidental Diamond Princess cruise ship exposures or community acquired COVID-19, with a diverse multinational population; and in Italy, CT scans varied from acute care screening to inpatients, commonly later in the disease process. Therefore, the conditions included in this study are fairly diverse, and we observed domain shifts among the three cohorts, as shown in Fig.~\ref{fig:ct}.

\noindent\textbf{Control Population} a control population was identified from one institution and one publicly available dataset: 1) 38 scans of 38 patients at the National Institutes of Health undergoing CT evaluation of known non-COVID-19 pneumonias from bacteria, fungi, and non-COVID viruses were included as a ``other pneumonia'' cohort (referred to as \textbf{Image\_P}), 2) 101 images of 101 patients with unremarkable lung findings from men with prostate cancer at the National Institutes of Health were included as a non-diseased ``normal'' cohort (referred to as \textbf{Image\_N}), and 3) a total of 474 scans of 474 patients were derived from the publicly available LIDC dataset~\citep{LIDC} consisting of lung nodule data (referred to as \textbf{Image\_LIDC}).

\noindent\textbf{Annotation} CT scans underwent a centralized evaluation by two expert radiologists for confirmation and localization of lung disease patterns related to COVID-19. Regions of CT infiltrates were manually delineated using ITK-SNAP tool~\citep{itksnap}. The most common ``CT infiltrate'' was ground glass opacities, followed by consolidation. To simulate the uneven distribution of annotation resource, out of the entire dataset, 671 (out of 736), 88 (out of 496), and 186 (out of 472) scans were annotated for Image\_1, Image\_2, and Image\_3, respectively.
Note that here, the three datasets have different degree of inter-observer variance from experts: all data of Image\_1 and Image\_2 are annotated by the same expert radiologists in the same institute, while Image\_3 is annotated by different radiologists from 2 different countries.

\begin{table*}[]
\centering
\begin{tabu}{cccccccc}
\toprule
&&\multicolumn{2}{c}{Image\_1}&\multicolumn{2}{c}{Image\_2}&\multicolumn{2}{c}{Image\_3}\\\cline{3-8}
&&Valid. Acc.&Test Acc.&Valid. Acc.&Test Acc.&Valid. Acc.&Test Acc.\\
\midrule
\multicolumn{2}{c}{Image\_1 (1 GPU)}&$0.571$\small$\pm 0.005$&$0.575$\small$\pm 0.003$&-&$0.578$\small$\pm 0.013$&-&$0.577$\small$\pm 0.011$\\
\multicolumn{2}{c}{Image\_2 (1 GPU)}&-&$0.479$\small$\pm 0.008$&$0.626$\small$\pm 0.004$&$0.625$\small$\pm 0.005$&-&$0.536$\small$\pm 0.010$\\
\multicolumn{2}{c}{Image\_3 (1 GPU)}&-&$0.480$\small$\pm 0.010$&-&$0.570$\small$\pm 0.015$&$0.649$\small$\pm 0.004$&$0.607$\small$\pm 0.007$\\
\multicolumn{2}{c}{Image\_1, Image\_2 (1 GPU)}&$0.572$\small$\pm 0.003$&$0.578$\small$\pm 0.006$&$0.634$\small$\pm 0.005$&$0.593$\small$\pm 0.011$&-&$0.579$\small$\pm 0.013$\\
\multicolumn{2}{c}{Image\_1, Image\_2 (2 GPUs)}&$0.581$\small$\pm 0.004$&$0.601$\small$\pm 0.004$&$0.639$\small$\pm 0.007$&$0.594$\small$\pm 0.010$&-&$0.575$\small$\pm 0.009$\\
%
% \rowfont{\color{red}}
\multicolumn{2}{c}{Image\_1, Image\_3 (1 GPU)}&$0.564$\small$\pm 0.003$&$0.574$\small$\pm 0.003$&-&$0.580$\small$\pm 0.017$&$0.636$\small$\pm 0.005$&$0.574$\small$\pm 0.015$\\
% \rowfont{\color{red}}
\multicolumn{2}{c}{Image\_1, Image\_3 (2 GPUs)}&$0.575$\small$\pm 0.008$&$0.586$\small$\pm 0.008$&-&$0.605$\small$\pm 0.001$&$0.651$\small$\pm 0.007$&$0.600$\small$\pm 0.008$\\
%
% \rowfont{\color{red}}
\multicolumn{2}{c}{Image\_2, Image\_3 (1 GPU)}&-&$0.489$\small$\pm 0.008$&$0.627$\small$\pm 0.006$&$0.600$\small$\pm 0.007$&$0.650$\small$\pm 0.008$&$0.615$\small$\pm 0.010$\\
% \rowfont{\color{red}}
\multicolumn{2}{c}{Image\_2, Image\_3 (2 GPUs)}&-&$0.513$\small$\pm 0.011$&$0.640$\small$\pm 0.003$&$0.613$\small$\pm 0.007$&$0.665$\small$\pm 0.004$&$0.638$\small$\pm 0.005$\\
%
% \rowfont{\color{red}}
\multicolumn{2}{c}{Image\_1, Image\_2, Image\_3 (1 GPU)}&$0.565$\small$\pm 0.010$&$0.572$\small$\pm 0.010$&$0.623$\small$\pm 0.006$&$0.579$\small$\pm 0.014$&$0.637$\small$\pm 0.008$&$0.575$\small$\pm 0.016$\\
% \rowfont{\color{red}}
\multicolumn{2}{c}{Image\_1, Image\_2, Image\_3 (2 GPUs)}&$0.577$\small$\pm 0.006$&$0.590$\small$\pm 0.008$&$0.642$\small$\pm 0.002$&$0.608$\small$\pm 0.008$&$0.647$\small$\pm 0.006$&$0.591$\small$\pm 0.008$\\
\midrule
FL Client 0&FL Client 1&&&&&\\\cline{1-8}
Sup., Image\_1&Sup., Image\_2&$0.566$\small$\pm 0.002$&$0.573$\small$\pm 0.005$&$0.637$\small$\pm 0.016$&$0.603$\small$\pm 0.019$&-&$0.579$\small$\pm 0.010$\\
Sup., Image\_1&Unsup., Image\_2&$0.563$\small$\pm 0.004$&$0.569$\small$\pm 0.006$&-&$0.590$\small$\pm 0.011$&-&$0.568$\small$\pm 0.011$\\
% \rowfont{\color{red}}
Unsup., Image\_1&Sup., Image\_2&-&$0.478$\small$\pm 0.007$&$0.628$\small$\pm 0.006$&$0.622$\small$\pm 0.004$&-&$0.553$\small$\pm 0.015$\\
%
% \rowfont{\color{red}}
Sup., Image\_1&Sup., Image\_3&$0.549$\small$\pm 0.010$&$0.552$\small$\pm 0.009$&-&$0.566$\small$\pm 0.023$&$0.628$\small$\pm 0.018$&$0.560$\small$\pm 0.028$\\
% \rowfont{\color{red}}
Sup., Image\_1&Unsup., Image\_3&$0.561$\small$\pm 0.003$&$0.564$\small$\pm 0.001$&-&$0.569$\small$\pm 0.010$&-&$0.551$\small$\pm 0.010$\\
% \rowfont{\color{red}}
Unsup., Image\_1&Sup., Image\_3&-&$0.458$\small$\pm 0.010$&-&$0.527$\small$\pm 0.032$&$0.637$\small$\pm 0.008$&$0.579$\small$\pm 0.023$\\
%
% \rowfont{\color{red}}
Sup., Image\_2&Sup., Image\_3&-&$0.489$\small$\pm 0.007$&$0.616$\small$\pm 0.011$&$0.588$\small$\pm 0.017$&$0.664$\small$\pm 0.004$&$0.605$\small$\pm 0.010$\\
% \rowfont{\color{red}}
Sup., Image\_2&Unsup., Image\_3&-&$0.471$\small$\pm 0.014$&$0.627$\small$\pm 0.004$&$0.625$\small$\pm 0.008$&-&$0.541$\small$\pm 0.017$\\
% \rowfont{\color{red}}
Unsup., Image\_2&Sup., Image\_3&-&$0.461$\small$\pm 0.004$&-&$0.551$\small$\pm 0.019$&$0.637$\small$\pm 0.009$&$0.586$\small$\pm 0.019$\\
\bottomrule
\end{tabu}
\caption{Accuracy (Dice's score) comparison with different experimental settings. The upper part of the table is for regular training with 1 and 2 GPUs, the bottom part is for federated learning with two clients. ``\textbf{Sup.}'' indicates the supervised client, and ``\textbf{Unsup.}'' indicates the client which contributes unlabeled data only.}
% and $*$ denotes the FL training is from scratch.}
\label{tab:res}
\end{table*}

\begin{figure*}[h]
\begin{center}
    \includegraphics[width=\linewidth]{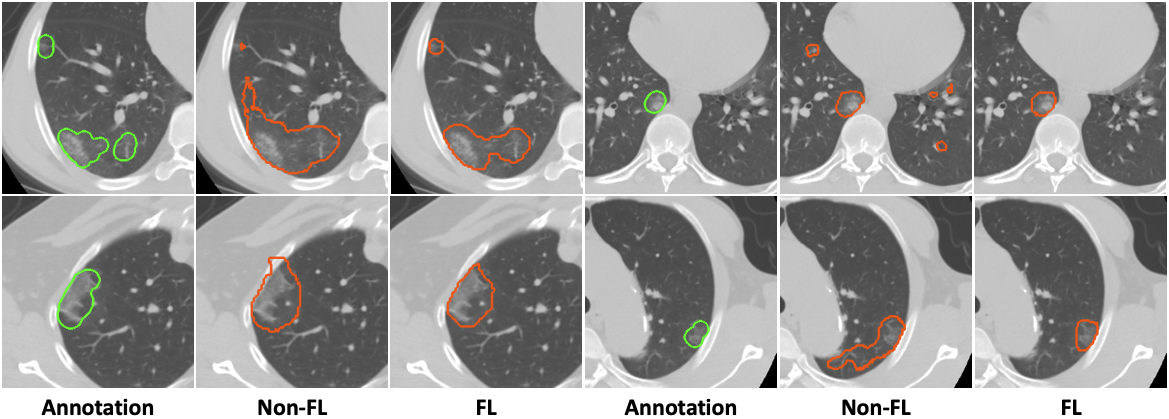}
\end{center}
%\vspace{-0.5cm}
\caption{
    Visualizations of federated semi-supervised segmentation of COVID regions in 3D CT (from the testing set of unsupervised client \textbf{Image\_2}). ``Non-FL'' indicates results from the model trained with Image\_1 along, and ``FL'' denotes results from the model trained with federated semi-supervised learning on Image\_1 and Image\_2. The segmentation results using the proposed framework captures the ground truth shapes better and has less false positives.
}
\label{fig:pred}
\end{figure*}

\subsection{Analysis}
\noindent\textbf{Data Split and Experiment Design} Among the images with experts' annotation, we split the dataset randomly into training/validation/testing sets. To ensure sufficient testing and validation data, the following splits are used: 447/112/112 for Image\_1, 30/29/29 for Image\_2, and 124/31/31 for Image\_3. To test the system's performance under an unbalanced situation, for most experiments, we use two of the three datasets as two clients for federated learning, while the other is set up as an ``unseen'' domain in order to test the models' generalizability.  

\noindent\textbf{Supervised Baselines} The baselines for our study is the model learnt under fully supervised conditions, this includes models trained with single source datasets, as well as centralized training with mixed Image\_1, Image\_2, and Image\_3 datasets. Here, three combinations of two sites are performed matching the experiments of federated learning, where we always want to keep an unseen domain so as to evaluate the generalizability of the trained model.

Further note that for fair comparison with federated learning where two sites train on their own GPU, besides training with 1 GPU, we also added a study with 2 GPU training. Also to reduce the influence from training randomness, each baseline is repeated five times under deterministic training with different random seeds, and the mean performance with standard deviation is reported.

\noindent\textbf{Parameter Setting}~The setting of clients was chosen empirically with some trials of different combinations and selecting the ones with the best performance as following: hard Dice loss on foreground with $\tau=0.90$, supervised client learning rate 1e-4, unsupervised client learning rate 5e-6. The augmentation for computing consistency loss is random scale shift of intensity.

As shown in Tab.~\ref{tab:res}, single-site training suffers from generalizability on other domains, models trained on Image\_2 and Image\_3 has~$10\%$ Dice drop for the testing accuracy on Image\_1 comparing to the model trained on Image\_1, and similar patterns can be observed from the other two scenarios. 

The performance of a model on a set of data depends on: 1) the capability and generalizability of the model, 2) the data distribution and similarity between the training and testing data, and 3) the annotation variation for evaluation. Therefore, ``unseen domain'' does not necessarily mean ``domain with significant shift'', and furthermore, ``domain with significant shift'' does not necessarily mean ``domain where the model will see significant performance drop''. Meanwhile, the ``performance'' also has two dimensions: 1) absolute performance with the same model on different datasets; and 2) relative performance on the same data with different models. Absolute performance is hard to predict, while in most cases, relative performance with model trained on the same data has the highest score. In other words, the following holds for most cases:
Two datasets 1 and 2, let's denote the performance of the model trained on i and tested on j as $p_{ij}$, then we have: for absolute performance $p_{11}$ may or may not $> p_{12}$, similarly, $p_{22}$ may or may not $>p_{21}$; on the other hand in most cases for relative performance, $p_{11} > p_{21}$, and $p_{22} > p_{12}$. This can be observed from Tab.~\ref{tab:res}: first row, absolute performance $p_{11}<p_{12}$; while fourth column, relative performance $p_{22}>p_{12}$.

Since the performance and robustness of a model is largely determined by the training data, if the supervised client is ``strong'', i.e. containing sufficient amount of data (in our case Image\_1), the trend of adding more data (Image\_2/3), whether supervised or unsupervised, is fairly stable, and we list the observations below.

Combining datasets in a centralized training can help promoting both the accuracy and the generalizability of a model.
The gap of data distributions from two datasets is mitigated via randomized mini-batching.
Thus, models trained with data centralization perform better than the models on Image\_1 or Image\_2 independently.
Since FL is also capable to collect information from multiple datasets, FL trained models with both supervised clients have comparable performance to single-GPU trained model on Image\_1 or Image\_2.

Federated semi-supervised learning has approximately $1\%$ Dice's score drop compared to supervised FL.
However, the model performs better than the model trained on Image\_1, Image\_2, or Image\_3 independently in general (see the testing results on Image\_2 and Image\_3).
It demonstrates that \textbf{the unlabeled data from different clients are valuable to train a generalizable model}.
The visual results are shown in Fig.~\ref{fig:pred}, and our model predicts segmentation masks with better shape and less false positives.

On the other hand, if the supervised client is not ``strong'' or less representative of overall data distribution, i.e. containing limited amount of data (in our case Image\_2), or containing annotation variance/bias (in our case ``Image\_3''), then the trend of adding more data can become unstable.
% To be specific, large data/annotation variances (e.g. dataset ``Image\_3'' compared to others) can introduce significant difficulties to fully- or semi-supervised federated learning (shown in Tab.~\ref{tab:res}).

First, federated learning with ``Image\_1'' and ``Image\_2'' meets our expectation using both fully- or semi-supervised setting. The supervised clients' models maintains the same level performance, and the model generalizability has been further improved over three datasets. Second, the dataset ``Image\_3'' seems to possess  bias regarding to overall data distribution. Adding ``Image\_3'' into training may down-grade the performance a bit (line ``Unsup., Image\_2 Sup., Image\_3'' in Tab.~\ref{tab:res}), which is potentially caused by biased annotation protocols from different radiologists of different countries. However, unsupervised ``Image\_3'' client (appearance of ``Image\_3'') is still able to help supervised client ``Image\_2'' to improve its model generalizability (line ``Sup., Image\_2 Unsup., Image\_3'' in Tab.~\ref{tab:res}). Moreover, due to much larger size of ``Image\_1'' compared to ``Image\_2'' and ``Image\_3'', unsupervised client ``Image\_1'' would down-grade its own performance by a large margin compared to its supervised counter-part.

Because federated learning is one type of distributed learning using multiple computing units.
The centralized 2-GPU training was conducted to make further comparison.
From Tab.~\ref{tab:res}, we can see the 2-GPU centralized training performs much better than other setting.
It is mainly caused by multi-GPU training with larger batch size per iteration.
The supervision from both dataset is stronger and pulls the model weights towards a stable condition.
\subsection{Ablation Studies}
Several components are configurable under the proposed federated semi-supervised learning framework, specifically, the image generation and loss functions for the unsupervised clients, the learning rate of each client, and the aggregation frequency for the server. The default FL framework followed the setting described above. We used same random seeds for all ablation study experiments. Due to the high amount of ablation studies, we trained a single model for each configuration instead of repeating 5 times. 

% random seeds five times
\noindent\textbf{Loss functions of self-/un-supervised learning}~In addition to foreground hard Dice loss, We experimented with another nine potential loss functions for $\mathcal{L}_\mathrm{consistency}$, including L1, L2, cross entropy (CE), and hard and soft Dice (Dice\_H and Dice\_S), all losses have two configurations: whole image and foreground (\_F). Tab.~\ref{tab:loss} summarizes their performance. As shown in the table, the hard Dice generally has an edge over other losses, while using foreground only may not yield better performance.  

\begin{table}[h]
\centering
\begin{tabu}{lcccc}
\toprule
&\multicolumn{2}{c}{Image\_1}&{Image\_2}&{Image\_3}\\
&\multicolumn{2}{c}{Sup.}&{Unsup.}&{Unseen}\\\cline{2-5}
&$\mathrm{Acc}_\mathrm{Valid.}$&$\mathrm{Acc}_\mathrm{Test}$&$\mathrm{Acc}_\mathrm{Test}$&$\mathrm{Acc}_\mathrm{Test}$\\
\midrule
Baseline&0.562&0.571&0.546&0.556\\
\midrule
\multicolumn{2}{l}{\textbf{Loss Function}}&&&\\
L1&0.540&0.552&0.585&0.571\\
L1\_F&0.553&0.548&0.552&0.547\\
L2&0.549&0.552&0.579&0.548\\
L2\_F&0.540&0.539&0.575&0.548\\
CE&0.557&0.560&0.593&0.564\\
CE\_F&0.559&0.571&0.535&0.545\\
Dice\_S&0.565&0.567&0.592&0.565\\
Dice\_SF&0.552&0.561&0.576&0.546\\
Dice\_H&\textbf{0.574}&0.569&\textbf{0.596}&0.569\\
\midrule
\multicolumn{2}{l}{\textbf{Data Augmentation}}&&&\\
Gauss\_0.1&0.550&0.552&0.557&0.566\\
Gauss\_0.9&0.548&0.551&0.578&0.560\\
Cutout\_1&0.548&0.548&0.569&0.543\\
Cutout\_2&0.547&0.556&0.555&0.529\\
Fix\_0.1&0.554&0.559&0.591&0.556\\
Fix\_0.25&0.552&0.549&0.572&0.547\\
\midrule
\multicolumn{2}{l}{\textbf{Aggregation Weights}}&&&\\
% \rowfont{\color{red}}
1.0:1.0&0.562&0.571&0.546&0.556\\
1.0:0.75&0.560&\textbf{0.573}&0.574&0.553\\
1.0:0.5&0.563&0.562&0.588&0.570\\
1.0:0.25&0.562&0.558&0.592&0.570\\
1.0:0.1&0.563&0.568&0.572&0.554\\
\midrule
\multicolumn{2}{l}{\textbf{Partial Weight Update}}&&&\\
Final&0.535&0.550&0.598&0.555\\
Block\_10&0.557&0.564&0.573&0.564\\
Block\_6&0.564&0.572&0.592&\textbf{0.578}\\
\bottomrule
\end{tabu}
\caption{Ablation studies for unsupervised loss, augmentation functions of unsupervised client, and aggregation weights.}
\label{tab:loss}
\end{table}

\noindent\textbf{Data augmentation}~We further conduct comparison against different augmentation strategies for unsupervised clients.
``Gauss\_$x$'' means adding voxel-wise Gaussian noise into CT volumes with zero mean and variance $x$.
From Tab.~\ref{tab:loss}, higher noise level increases generalizability of the model globally.
``Cutout\_1''~\cite{devries2017improved} means masking a random cube (size $20 \times 20 \times 3$) from the re-sampled image with zero, and ``Cutout\_2'' means masking five of such random cubes from the re-sampled image with zero.
The performance drops significantly when increasing the ``Cutout'' regions in augmentation (shown in results on Image\_3).
``Fix\_$y$'' follows the idea of ``FixMatch''~\citep{sohn2020fixmatch} to create both strongly and weakly augmented samples for unsupervised learning.
And the pseudo-label is generated based on the weakly augmented samples.
$y$ means the level of random intensity shift for weakly augmented samples, and $1.0 - y$ is for strongly augmented samples.
$y=0.1$ creates larger gap between strongly and weakly augmented samples, compared to $y=0.25$.
Clearly from the Tab.~\ref{tab:loss}, the larger difference of samples corresponds to the more generalizable models. As for a particular task and data, experiments will be needed to select the best data augmentation strategies for computing consistency loss.

\noindent\textbf{Aggregation weights}~During each round of server aggregation, updates from different clients can be weighted before they are aggregated together to update the model on the server end. Here, we adjust the weight of the unsupervised client from 0.1 to 1.0. As shown in Tab.~\ref{tab:loss}, the behavior of this parameter is not linearly correlated with the performance on specific datasets. This may be explained by the fact that the updates learnt from unsupervised client, while catching characteristics of Image\_2, also contains fair amount of noise due to its unsupervised nature. Therefore, reducing the weight from unsupervised client leads to not only the reduction of the influence from Image\_2, but also the noise from it.  

\noindent\textbf{Partial model update}~Another interesting experiment is to share partial weights only across clients.
In Tab.~\ref{tab:loss}, ``final'' denotes the last convolutional layers are not shared between clients, ``block\_10'' denotes the last DenseBlock and final convolutoinal layers together are not shared, and ``block\_6'' illustrates all layers after encoder (equivalent to decoder path plus last convolutional layer, blocks 6, 7, 8, 9, 10 $+$ last convolutional layer) are not shared.
From the comparison, the less layers are shared, the better performance the global model has.
The FL model, with encoder-only sharing, possess excellent performance with better generalizability for both seen and unseen datasets.
And such finding might be caused by intrinsic difference between supervised and unsupervised tasks.
Federated encoder training could jointly learn a good feature representation across multiple clients' database, then the segmentation task is handled better with independent decoder with labeled data.
Meanwhile, the similar strategy could be applied even for FL with all supervised clients, or multi-task FL.
Furthermore, the partial weight sharing provides safer solutions for privacy preserving, since only partial model information is used in client-server communication.
The experimental results in the paper, other than ones in this ablation study, are federated averaging with the full models (aggregating all model weights).

\begin{table}[h]
\centering
\begin{tabular}{ccccc}
\toprule
&\multicolumn{2}{c}{Image\_1}&{Image\_2}&{Image\_3}\\
&\multicolumn{2}{c}{Sup.}&{Unseen}&{Unseen}\\\cline{2-5}
&$\mathrm{Acc}_\mathrm{Valid.}$&$\mathrm{Acc}_\mathrm{Test}$&$\mathrm{Acc}_\mathrm{Test}$&$\mathrm{Acc}_\mathrm{Test}$\\
\midrule
N&$0.548$&$0.546$&$0.588$&$0.550$\\
LIDC&$0.550$&$0.564$&$0.593$&$0.562$\\
P&$0.566$&$0.562$&$0.591$&$0.556$\\
\midrule
\midrule
SegResNet&0.533$\pm$&0.540$\pm$&0.596$\pm$&0.489$\pm$ \\
Image\_1&0.002&0.004&0.003&0.006\\ 
\midrule
SegResNet&0.534&0.543&0.607&0.512\\
\bottomrule
\end{tabular}
\caption{Ablation studies for unsupervised client datasets and another base network.}
\label{tab:data}
\end{table}

\noindent\textbf{Other datasets and other network}
The control population of Image\_N, Image\_LIDC, and Image\_P can also serve for the unsupervised client. From the cohort relationship perspective, Image\_N is farthest from the candidate COVID-19 data, since it contains almost no abnormal regions. Image\_LIDC may have certain similarity, since some COVID-19 cases can have nodule-like focal consolidation areas. Image\_P would be the most close to COVID-19 dataset, since COVID is common for scans of pneumonia. As illustrated from the results listed in Tab.~\ref{tab:data}, the above relationship can be observed from the validation accuracy, which is the most predicable. The testing accuracy for Image\_1 also shows a similar pattern, though Image\_LIDC and Image\_P have similar performance. The testing accuracy on the two unseen domains are more unpredictable. 
Besides the current base network~\citep{liu20183d}, other networks serving similar purpose can also be used. We did an experiment with the SegResNet proposed in~\citep{segresnet}.  The single site training on Image\_1 is also trained five times, and the federated semi-supervised learning is setup with the same configurations of loss, etc. as previous baseline study. As shown in Tab.~\ref{tab:data}, the overall accuracy is not as high as the current network, but the federated result is slightly better than Image\_1 alone. Again as mentioned previously, our framework is flexible to host most networks as the base network. 

\begin{figure}[h]
\begin{center}
    \includegraphics[width=9cm]{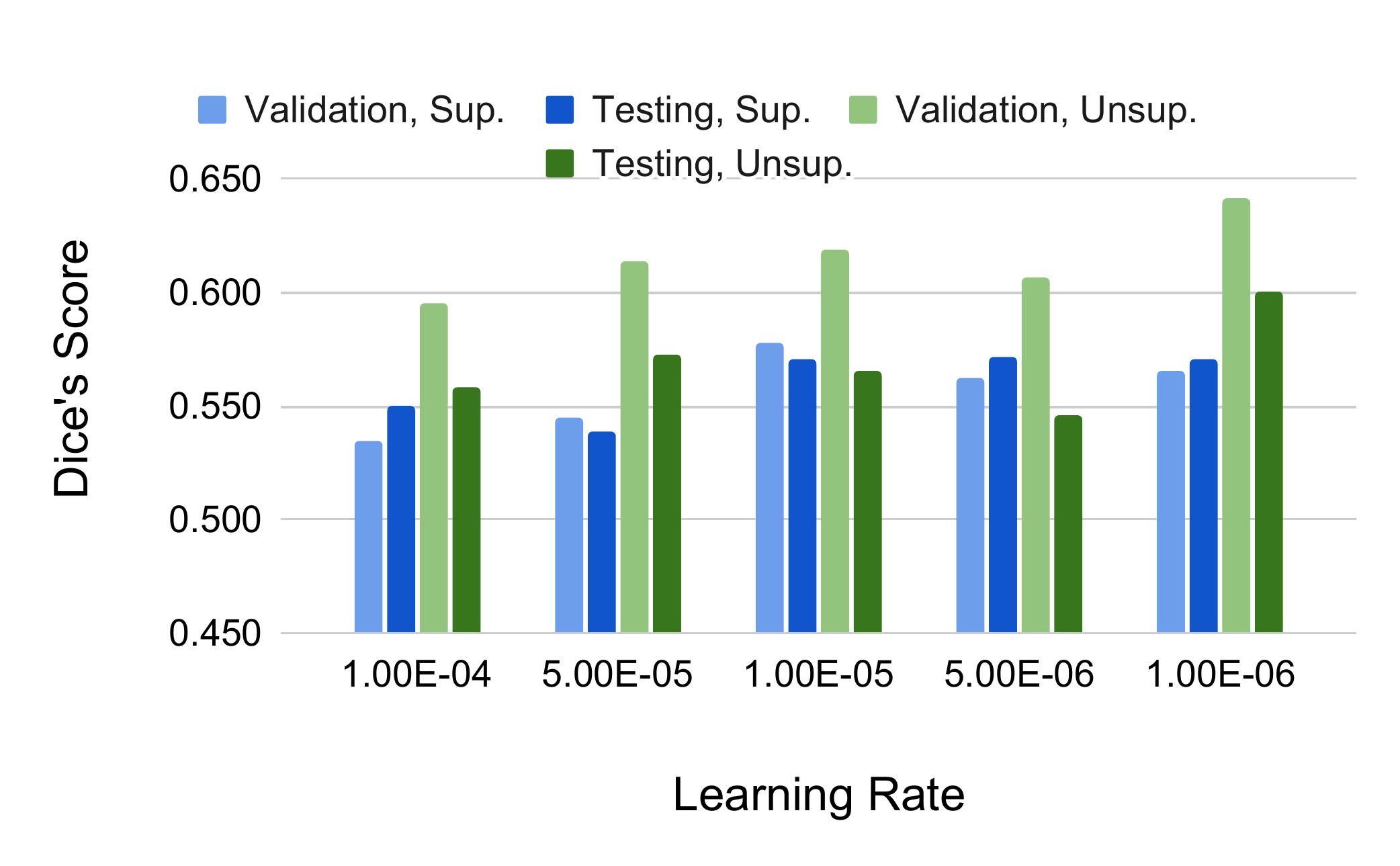}
\end{center}
\caption{
\footnotesize Accuracy (Dice's score) comparison of different learning rates of un-/unsupervised clients.
}
\label{fig:lr}
\end{figure}

\noindent\textbf{Learning rates}~We studied different learning rates (LR) on the unsupervised client, while LR of the supervised client is fixed at 1e-4 for fair comparison. The value of LR varies from 1e-6 to 1e-4. And LR of the unsupervised clients cannot be too large or even larger than the one used in the supervised clients.
In the FL setting, each round of training is client-independent.
Therefore, the model weights would quickly converge to over-fit the self-supervised tasks when LR is large.
It indicates that the over-fitted model weights might be biased towards unexpected directions, which would make the overall training procedure unstable.
Such side-effect is shown in the Fig.~\ref{fig:lr}.
Here, the supervised client is using Image\_1, and unsupervised client is using Image\_2.
The accuracy on the validation/testing set of unsupervised client is lower, when LR is set to a higher values.
Higher LR also affects the performance on the supervised client, which demonstrates that 1) FL training become less effective; 2) the correlation between self-supervised tasks and supervised tasks is weak.
Thus, lower LR on unsupervised client generally benefits all clients in FL.

\begin{figure}[h]
\begin{center}
    \includegraphics[width=9cm]{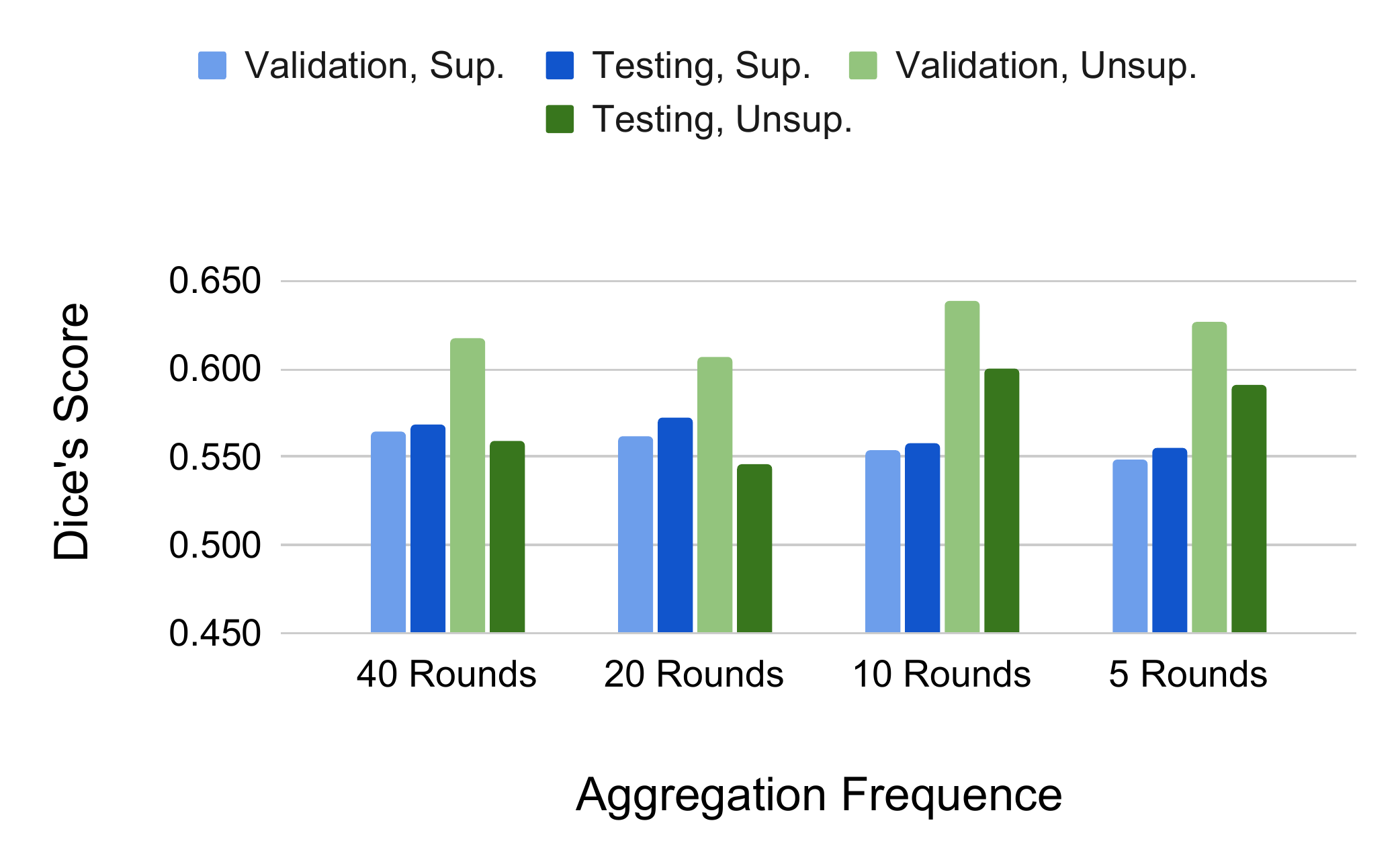}
\end{center}
\caption{
\footnotesize Accuracy (Dice's score) comparison of different aggregation frequency (per 5, 10, 20, 40 rounds).
}
\label{fig:af}
\end{figure}

\noindent\textbf{Aggregation frequency}~Another question for FL is how often clients and server should communicate to each other.
In other words, how often should client model weights be aggregated.
Ideally, if the model aggregation happens every training iteration, then federated semi-supervised learning is equivalent with standard semi-supervised learning with equal sampling chance from both labeled and unlabeled datasets.
In order to verify the effects of aggregation frequency, we conducted the ablation study with aggregation per 5, 10, 20, 40 epochs shown in Fig.~\ref{fig:af}.
Here, supervised client is using Image\_1, and unsupervised client is using Image\_2.
In general, when the aggregation is more frequent, the performance on the self-supervised side improves (see the performance on the testing data of the unsupervised client).
At the same time, the performance of the supervised clients become slightly worse.
More frequent aggregations correct the training trajectory on unsupervised clients more often, and the overall training becomes smoother.
In reality, it is not practical to synchronize clients frequently due to the bandwidth limitations. The trade-off between the aggregation frequency and the communication cost needs to be tuned for the optimal training efficiency depending on the specific applications.

% \noindent\textbf{Aggregation weights}~
% \noindent\textbf{Partial model update}~

\section{Conclusions and Discussion}
\begin{figure}[h]
\begin{center}
    \includegraphics[width=9cm]{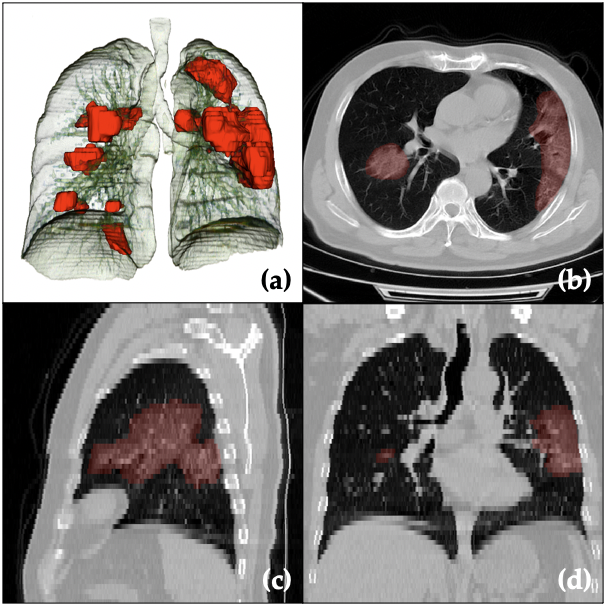}
\end{center}
\caption{
    \footnotesize
    % \revision{
        Visualization of the COVID-19 affected region segmentation/prediction (red regions) together with lungs and airways (a) in 3D space, and (b,c,d) in different (axial, sagittal, coronal) planes of a raw CT image. Note that slice thickness is 5 mm (as compared with 0.8 mm in-plane), which is the case for most images in this work.
    % }
}
\label{fig:vis2}
\end{figure}

In this paper, we proposed a \textit{federated semi-supervised learning} framework for COVID-19 affected region segmentation in 3D chest CT (3D visualization with airway and lung in Fig~\ref{fig:vis2}).
The proposed framework is capable to grasp valuable information from the clients which only have unlabeled data.
Meanwhile, the privacy of all patients has been preserved, and they do not need to share their own database for collaborating on joint model training.
Moreover, after jointly training with supervised and un-/unsupervised clients, the generalizability has been improved for not only each client's database, but also on the unseen data domain.
We found out even the client with pure non-COVID database is able to help model training for COVID-19 affected region segmentation via false alarm rejection.

\begin{figure}[h]
\begin{center}
    \includegraphics[width=9cm]{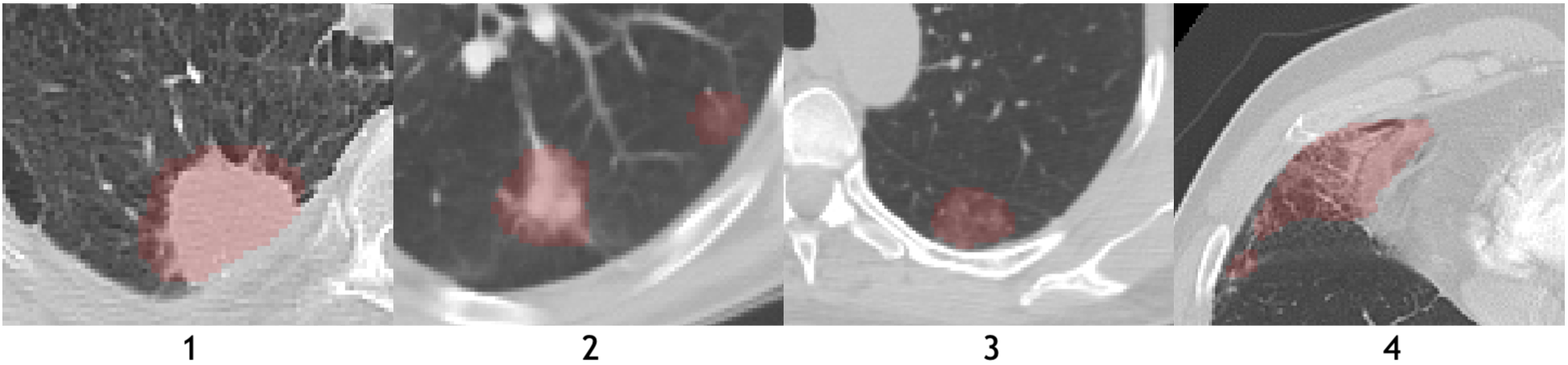}
\end{center}
\caption{
\footnotesize Model performance on LIDC dataset, capturing solid nodule (1, 2); ground glass nodule (2, 3); and other abnormal patterns (4).
}
\label{fig:lidc}
\end{figure}

One thing to further clarify is that the aim of work is to segment the disease affected regions that is reflected in CT images, and the annotation is solely based on dataset consisting of COVID-19 cases. The ``COVID-19 affected region'' is identified as the abnormal regions in the context of these cases. Therefore, the model is not trained to discriminate against other type of abnormalities, e.g. other pneumonia or cancer. From a ``pipeline'' point of view, additional classification to tell the difference can follow the proposed segmentation method, but will need additional data and annotations. To give an example, we used our trained network to perform inference on LIDC dataset~\cite{LIDC}. Fig.~\ref{fig:lidc} showed four different cases of abnormalities captured by our model on LIDC images: 1. solid nodule, 2. mixed solid and ground glass nodule, 3. ground glass nodule, and 4. other abnormal pattern. As nodules have similar appearance in CT as abnormalities caused by COVID-19, those regions are detected.   

Our proposed framework has been validated on a multi-national database cohort with population, equipment, and demographic variation.
In addition, comprehensive ablation studies have been explored for the proposed framework.

The proposed \textit{federated semi-supervised learning} framework is general for machine learning based applications of medical image analysis.
Given the limited literature, this work may initiate a promising direction for the future study of medical image analysis.
Still, there are open questions along this research direction.
For example, there are potential domain gaps between supervised and unsupervised clients.
It is an unsolved problem of how to better model this domain gap and mitigate it during federated learning.
Another example is how to adaptively aggregate contributions from different clients based on the quality and and not just the quantity of a clients' database.
Because there are a lot variables in the semi-supervised framework, and complexity is even higher compared to regular FL, we hope our work is a good starting point for future exploration.

% \section*{Acknowledgment}
% \label{sec:acknowledgment}

\bibliographystyle{model2-names.bst}
\biboptions{authoryear}
\bibliography{refs}

\end{document}